\documentclass[12pt,preprint]{aastex}

\usepackage[]{natbib}
\usepackage{graphicx}
\usepackage{amssymb}
\usepackage{color}
\usepackage{ulem}

\newcommand{\etacar}{$\eta$~Car}

\input{macro.inp}

\slugcomment{ver.~arXiv}

\shorttitle{Eta Carinae's Thermal Tail}
\shortauthors{Hamaguchi et al.}

\begin{document}

\title{Eta Carinae's Thermal X-ray Tail \\
Measured with \XMM\ and \NUS}

\author{
Kenji Hamaguchi\altaffilmark{1,2}, 
Michael F. Corcoran\altaffilmark{1,3},
Theodore R. Gull\altaffilmark{4}, 
Hiromitsu Takahashi\altaffilmark{5},
Brian Grefenstette\altaffilmark{6},
Takayuki Yuasa\altaffilmark{7},
Martin Stuhlinger\altaffilmark{8},
Christopher M. P. Russell\altaffilmark{4,9},
Anthony F. J. Moffat\altaffilmark{10},
Neetika Sharma\altaffilmark{2},
Thomas I. Madura\altaffilmark{1,3},
Noel D. Richardson\altaffilmark{10},
Jose Groh\altaffilmark{11},
Julian M. Pittard\altaffilmark{12},
Stan Owocki\altaffilmark{13}
}

\altaffiltext{1}{CRESST and X-ray Astrophysics Laboratory NASA/GSFC, Greenbelt, MD 20771}
\altaffiltext{2}{Department of Physics, University of Maryland, Baltimore County, 1000 Hilltop Circle, Baltimore, MD 21250}
\altaffiltext{3}{Universities Space Research Association, 7178 Columbia Gateway Dr., Columbia, MD 21044}
\altaffiltext{4}{Astrophysics Science Division, NASA Goddard Space Flight Center, Greenbelt, MD 20771}
\altaffiltext{5}{Department of Physical Sciences, Hiroshima University, Higashi-Hiroshima, Hiroshima 739-8526, Japan}
\altaffiltext{6}{Space Radiation Lab, California Institute of Technology, Pasadena, CA 91125}
\altaffiltext{7}{Nishina Center, RIKEN, 2-1, Hirosawa, Wako, Saitama, Japan, 351-0198, Japan}
\altaffiltext{8}{European Space Astronomy Centre (ESAC), P.O. Box 78, 28691 Villanueva de la Ca–ada, Madrid, Spain}
\altaffiltext{9}{NASA Postdoctoral Program Fellow}
\altaffiltext{10}{D\'epartement de physique and Centre de Recherche en Astrophysique du Qu\'ebec (CRAQ), Universit\'e de Montr\'eal, C.P. 6128}
\altaffiltext{11}{Geneva Observatory, Geneva University, Chemin des Maillettes 51, CH-1290 Sauverny, Switzerland}
\altaffiltext{12}{School of Physics and Astronomy, The University of Leeds, Woodhouse Lane, Leeds LS2 9JT, UK}
\altaffiltext{13}{Bartol Research Institute, Department of Physics and Astronomy, University of Delaware, Newark, DE 19716, USA}

\begin{abstract}
The evolved, massive highly eccentric binary system, \etacar, underwent a periastron passage in the summer of 2014.
We obtained two coordinated X-ray observations with \XMM\ and \NUS\ during the elevated X-ray flux state and 
just before the X-ray minimum flux state around this passage.
These \NUS\ observations clearly detected X-ray emission associated with \etacar\ extending up to $\sim$50~keV for the first time.
The \NUS\ spectrum above 10 keV can be fit with the bremsstrahlung tail from a \KT~$\sim$6 keV plasma. 
This temperature is $\Delta$\KT~$\sim$2~keV higher than those measured from the iron K emission line complex,
if the shocked gas is in collisional ionization equilibrium.
This result may suggest that the companion star's pre-shock wind velocity is underestimated.
The \NUS\ observation near the X-ray minimum state showed a gradual decline in the X-ray emission 
by 40\% at energies above 5 keV in a day, the largest rate of change of the X-ray flux yet observed in 
individual \etacar\ observations.
The column density to the hardest emission component, \NH~$\sim$10$^{24}$~\UNITNH, 
marked one of the highest values ever observed for $\eta$ Car, 
strongly suggesting increased obscuration of the wind-wind colliding X-ray emission by the thick primary 
stellar wind prior to superior conjunction.
Neither observation detected the power-law component in the extremely hard band 
that \INTEGRAL\ and \SUZAKU\ observed prior to 2011.
If the non-detection by \NUS\ is caused by absorption,
the power-law source must be small and located very near the WWC apex.
Alternatively, it may be that the power-law source is not related to either \etacar\ or the GeV $\gamma$-ray source.
\end{abstract}
\keywords{Stars: individual (\etacar) --- stars: early-type --- stars: winds, outflows ---  
binaries: general --- X-rays: stars}

\section{Introduction}
Massive binary systems drive shock plasma heating via the collision of winds from two stars (wind-wind collision: WWC).
With typical (pre-shock) wind speeds of $\ge 1000$~\UNITVEL, temperatures can reach as high as several tens of millions of Kelvin.
X-ray emission from these stable shocks provides important tests of shock physics,
and multiple X-ray observations of such systems have been performed for decades \citep[e.g.,][]{Corcoran2001a,Skinner2001,Pollock2005,Zhekov2010}.
While the spectrum below 10 keV is complicated by discrete line emission and absorption components,
the X-ray spectrum above 10~keV is relatively simple.
This high-energy emission therefore provides important clues on the condition of the maximum thermalized plasma where 
the winds collide head-on, while also providing important information about particle acceleration through the shock. 
This information also helps us understand the wind and stellar properties, which can be difficult to constrain from optical or 
UV observations for stars that are heavily obscured by interstellar and circumstellar matter.

Eta Carinae \citep[$d \sim$2.3~kpc,][]{Smith2006b} is one of the most massive stars in our Galaxy with an initial mass of 
$\gtrsim$100~\UNITSOLARMASS~\citep{Hillier2001}.
After the giant eruption of the 1840s, the star exhibited extreme mass loss indicating that it may be near the end of its lifetime.
The star itself cannot be seen directly at most wavelengths due to an optically-thick stellar wind \citep[\Mdot~$\sim$8.5$\times$10$^{-4}$\UNITSOLARMASS~yr$^{-1}$,][]{Groh2012},
but periodic variations over nearly all wavelength bands revealed the presence of a binary system,
with a highly eccentric ($e \sim$0.9) 5.54 year orbit \citep{Damineli1997,Corcoran2005,Damineli2008}.
The collision of the wind from the more luminous primary and the secondary star produces plasma 
that provides a luminous source of X-rays in the system. 
Since the primary star drives a dense, slow \citep[$V~\sim420$~\UNITVEL,][]{Groh2012} wind, the companion must have a
very fast wind of $\sim$3000~\UNITVEL\ in order for the WWC to produce the observed hot X-ray plasmas 
\citep{Pittard2002}.
The unseen companion should be, therefore, a massive O star or a Wolf-Rayet star \citep{VernerE2005a,Parkin2009,Mehner2010b}.

\begin{figure}[t]
\plotone{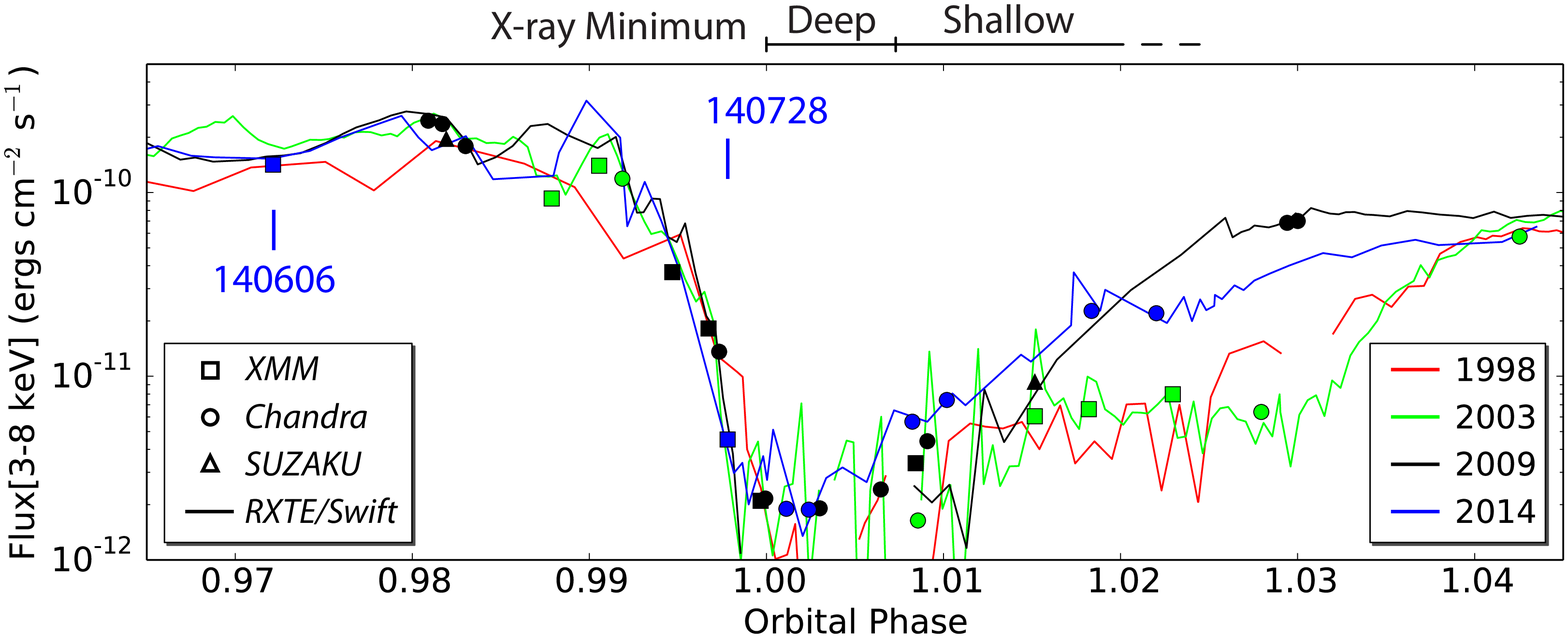}
\caption{
\RXTE\ and \SWIFT\ light curves of \etacar\ (Corcoran et al. 2015, in prep.) and 
the pointed observations \citep[][Hamaguchi et al., in prep.]{Hamaguchi2014a}.
The designations, 140606 and 140728, are timings of the coordinated observations of \XMM\ and \NUS.
The horizontal axis shows the orbital phase defined by \citet{Corcoran2005}.
The phase 1.0 corresponds to 2014 August 2 7:00:29 UT in this cycle.
}
\label{fig:variation_minimum}
\end{figure}

The WWC X-ray emission has been monitored intensively for 4 orbital cycles since 1996 \citep[][Corcoran et al., 2015 in prep.]{Corcoran2010}.
In every cycle, the observed X-ray emission increased dramatically by a factor of 3 toward periastron,
then suddenly declined to a minimum for a few months.
This X-ray minimum has two distinct phases \citep[][see Figure~\ref{fig:variation_minimum}]{Hamaguchi2007b}.
The first ``deep X-ray minimum" phase lasts approximately 3 weeks.
During this time, the WWC X-ray emission totally disappears
and residual emission from the central point source --- Central Constant Emission: CCE, \citep{Hamaguchi2007b,Hamaguchi2014a} ---
plus reflection of the WWC X-ray emission at the surrounding bipolar nebula
--- X-ray Homunculus Nebula: XHN \citep{Corcoran2004} --- 
is observed between 1$-$10~keV.
The following ``shallow X-ray minimum" is defined by a three-fold increase in X-ray emission.
It has been suggested that 
the deep minimum is produced by an eclipse of the WWC X-ray plasma by the optically thick primary wind,
while the shallow minimum is produced by the residual X-ray activity across periastron.

Extremely high energy X-rays near \etacar\ have been observed previously.
The \INTEGRAL\ observatory detected a point-like source around \etacar\ in the 22$-$100~keV band
in four pointed observations between 0.0$\lesssim \phi_{\rm orb} \lesssim$0.4
 \citep{Leyder2008,Leyder2010}.
The \SUZAKU\ observatory confirmed the presence of extremely high energy radiation in the 15$-$40~keV band
from the direction of \etacar\ \citep{Sekiguchi2009}.
Since no apparent high energy source other than \etacar\ has been found within the 2.4\ARCMIN\ \INTEGRAL\ error circle \citep{Leyder2010},
\etacar\ has been considered as the best candidate of the counterpart.
This emission did not vary remarkably throughout an entire single orbital cycle between 2005$-$2011,
suggesting little connection to the WWC thermal X-ray activity \citep{Hamaguchi2014b}.

These extremely hard X-rays are suspected to originate from the $\gamma$-ray source in the 0.1$-$100~GeV band near \etacar,
which was discovered by the \AGILE\ and \FERMI\ $\gamma$-ray observatories \citep{Tavani2009,Abdo2010}.
Again, \etacar\ is the only known high energy source within the error circle,
while the emission apparently varies slowly with the {\etacar}'s orbital period \citep{Reitberger2015}.
The spectrum shows two components, which may originate from
stellar UV photons up-scattered by Compton recoil of GeV electrons
that are accelerated by the 1st-order $Fermi$ mechanism at the WWC shocks,
or pion decay of TeV protons accelerated by the same mechanism and collided with surrounding wind material,
or both \citep{Abdo2010,Farnier2011,Ohm2015}.
This source was not detected in the very high-energy $\gamma$-ray (470~GeV$-$9~TeV) band with the HESS observatory,
suggestive of a spectral cut-off below 1~TeV \citep{HESSCollaboration2012a}.

In this paper, we present two joint broadband X-ray observations of \etacar\ with \XMM\ and \NUS\ at key orbital phases around periastron,
prior to the start of the deep X-ray minimum.
\XMM\ can obtain moderate resolution X-ray spectra below 10 keV including key spectral diagnostics like the Fe K emission line complex 
and the absorption structure of the Fe K edge, while \NUS\ can obtain direct imaging spectra in the hard X-ray band extending beyond 10 keV.
Because \NUS\ is the first focusing X-ray telescope above 10~keV, 
it also allows us to determine a more accurate location of the extremely hard X-ray source.
Using these observations,
we address some of the fundamental questions about the origin of the hard X-ray emission from \etacar.

\section{Observations}

In the summer of 2014, we observed \etacar\ with \XMM\ and \NUS\ simultaneously at two epochs around periastron (Table~\ref{tbl:obslogs}).
The first observation started on June 6 when \etacar\ was about to reach the X-ray maximum (Figure~\ref{fig:variation_minimum}).
The X-ray flux had already increased by a factor of 4 relative to the fluxes around apastron.
The second observation started on July 28 when the X-ray emission had dropped nearly two orders of magnitude from the X-ray maximum,
4 days before the beginning of the deep minimum phase, August 1,
according to monitoring observations by the X-ray Telescope on \SWIFT\ (Corcoran et al., in preparation).
For each observation, the \XMM\ observation covered only a part of the \NUS\ observation.
The \XMM\ observations were performed continuously,
while the \NUS\ observations were interrupted every $\sim$90 minutes by Earth occultation.
Following \citet{Hamaguchi2007b},
individual \XMM/\NUS\ observations are designated XMM/NUS,
subscripted with the year, month and day of the observation.

\XMM\ has three nested Wolter I-type X-ray telescopes \citep{Aschenbach2000} with 
the European Photon Imaging Camera (EPIC) CCD detectors
(pn, MOS1 and MOS2) in their focal planes \citep[][]{Struder2001, Turner2001}.
They achieve a spatial resolution of 17\ARCSEC\ half energy width
and an energy resolution of 150~eV at 6.4~keV\footnote{http://xmm.esac.esa.int/external/xmm\_user\_support/documentation/uhb/index.html}.
In each observation, \etacar\ was placed on-axis.
The EPIC-pn and MOS1 observations were obtained in the small window mode with the thick filter to avoid photon pile-up and optical leakage,
though the EPIC-MOS1 data in XMM$_{140606}$ was still affected by photon pile-up.
The EPIC-MOS2 observations used the full window mode with the medium filter to monitor serendipitous sources around \etacar,
so that its \etacar\ data are significantly affected by photon pile-up and optical leakage and 
thus provide no useful information about \etacar.
Fortunately, most of the \XMM\ observations were obtained during periods of low particle background.

\NUS\ has two nested Wolter I-type X-ray telescopes with a 2$\times$2 array of CdZnTe pixel detectors
in each focal plane \citep[FPMA/FPMB, ][]{Harrison2013}.
These mirrors are coated with depth-graded multilayer structures and focus X-rays over a 3$-$79 keV bandpass.
They achieve an angular resolution of roughly 60\ARCSEC\ half power diameter \citep{Madsen2015a}.
The focal plane detectors are sensitive above 3~keV and cover a 12\ARCMIN\ \FOV.
The energy resolution of the detectors is 400 eV below $\sim$40 keV, rising to $\sim$ 1 keV at 60 keV.
In each observation, \etacar\ was placed on-axis.
Because there are no bright sources ($>$100~mCrab) within 1\DEGREE\ to 5\DEGREE, stray light contamination was not an issue.

We used the analysis package HEASoft\footnote{http://heasarc.gsfc.nasa.gov/docs/software/lheasoft/}
version 6.16 and 6.17 and the SAS\footnote{http://xmm.esac.esa.int/sas/} version 14.0.0 and
Current Calibration Files (CCFs) as of 2014 December 9
for the \XMM\ specific data analysis. We used the \NUS\ calibration version 2015 March 20.

\section{X-ray Images}

\begin{figure}[t]
\plotone{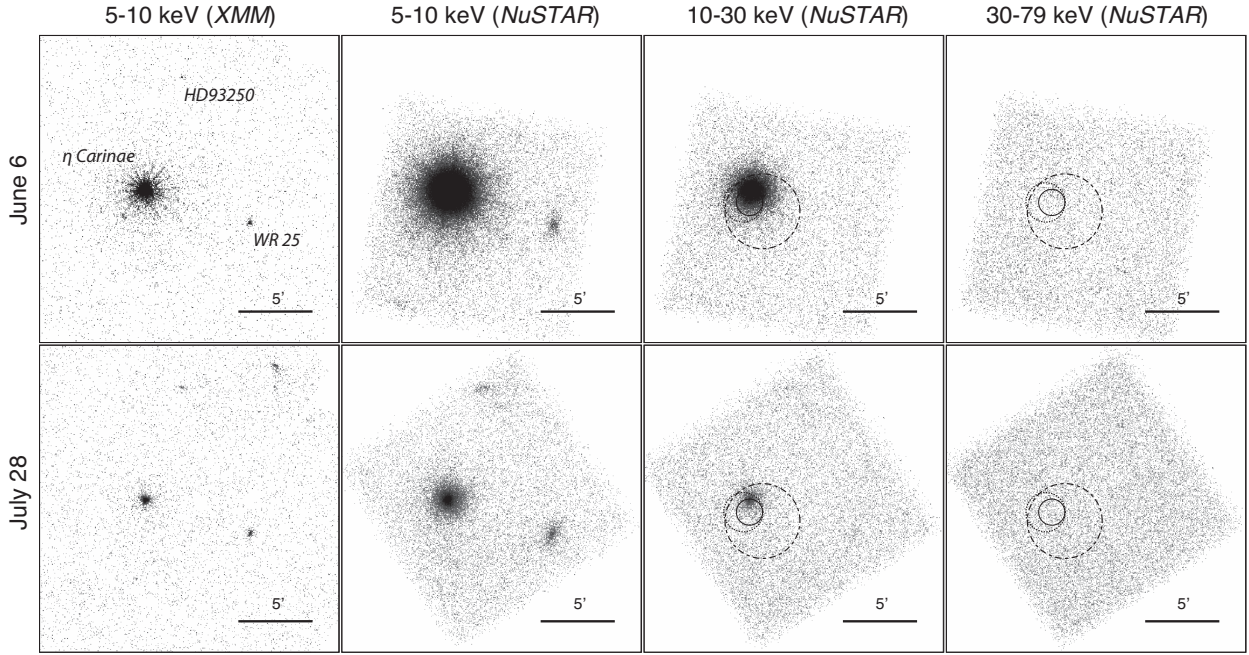}
\caption{
\XMM/EPIC-MOS2 (5$-$10~keV) and \NUS/FPMA$+$FPMB (5$-$10~keV, 10$-$30~keV and 30$-$79 keV) images of the \etacar\ field
during the first ({\it top}) and second ({\it bottom}) observations.
The grey scales of all images are adjusted with the event count rate.
In the images in the right two columns, 
the dashed-dot bar circles show the 90\% confidence range of the \INTEGRAL\ source \citep{Leyder2010}
and the solid and dotted circles the 95.4\% confidence ranges of the \FERMI\ source in the low-energy and high-energy
bands, respectively \citep{Reitberger2015}.
The EPIC-MOS2 data were not used for the timing and spectral analysis 
because the \etacar\ data suffered severe pile-ups.
\label{fig:image}
}
\end{figure}

Figure~\ref{fig:image} shows the \XMM\ EPIC-MOS2 (5$-$10 keV) and
the \NUS\ FPMA+FPMB (5$-$10~keV, 10$-$30~keV, 30$-$79~keV) images of each observation.
These \NUS\ images are the first images of the Carina Nebula near \etacar\ at $E>$10~keV at this spatial resolution ($\sim$1\ARCMIN).
Eta Carinae at the \FOV\ center is the brightest source below 30 keV;
the source position does not shift significantly between the energy bands.
In the 30$-$79~keV band, \etacar\ is barely seen in NUS$_{140606}$ and not at all in NUS$_{140728}$.
There are no other X-ray point sources detected at energies above 10~keV within the error circles of the \FERMI\ 
and \INTEGRAL\ source positions, which are shown by circles in the two right column images of Figure~\ref{fig:image}.
The images below 30~keV also show the WWC binary system, WR25, and the massive O star HD~93250.

\section{Light Curves and Spectra}

\subsection{Event Extraction and Estimate of the Stable Component}
\label{subsec:event_extract}

We followed \citet{Hamaguchi2007b} for extracting  \XMM\ source light curves and spectra,
taking the \etacar\ source region from a 50\ARCSEC$\times$37.5\ARCSEC\ ellipse 
with the major axis rotated from the west to the north at 30\DEGREE.
For background estimation, we used regions with negligible emission from \etacar\ on the same CCD chip.
In addition, we limited the EPIC-pn background regions at around the same RAWY position of \etacar,
according to the \XMM\ analysis guide\footnote{http://xmm.esac.esa.int/sas/current/documentation/threads/PN\_spectrum\_thread.shtml}.

We extracted \NUS\ source events from a 50.5\ARCSEC\ radius circle centered on \etacar,
which includes 70\% of photons from the star\footnote{http://www.nustar.caltech.edu/uploads/files/nustar\_performance\_v1.pdf}.
Though this source region is slightly larger than the \XMM\ source region,
hard X-ray ($\gtrsim$2~keV) emission from \etacar\ is constrained to within $\sim$10\ARCSEC\ from the star \citep{Hamaguchi2014a},
so that the small discrepancy in the \XMM\ and \NUS\ source regions should not be significant.
For the \NUS\ observations, we extracted backgrounds from a 630\ARCSEC\ squared box region inside the detector \FOV,
excluding the region within 200\ARCSEC\ or 300\ARCSEC\ from \etacar\ and 128\ARCSEC\ of the other X-ray sources detected with \NUS.
We extracted light curves and spectra using the HEASoft tool, {\tt nuproduct}.

In addition to the WWC X-rays, \etacar\ shows weak, stable CCE emission and time-delayed XHN emission,
which make a non-negligible contribution to the \etacar\ spectra near X-ray minimum \citep[see][]{Hamaguchi2014b}.
We estimated the contribution of these components using a \SUZAKU\ observation, which we obtained on 2014 August 6 
during the deep minimum (ObsID: 409028010).
We extracted spectra from the \SUZAKU\ XIS0, 1 and 3 detectors from a circular region of radius 2.5\ARCMIN\ centered on 
the source and fit these spectra by a 2-temperature plasma ({\tt apec}) components with individual absorption components,
including two Gaussians for the fluorescent Fe K$\alpha$ and K$\beta$ lines.
We scaled the XIS1 and XIS3 model normalization to 1.026 and 1.014, respectively, of the XIS0 normalization,
following the \SUZAKU\ data analysis
guide\footnote{http://heasarc.gsfc.nasa.gov/docs/suzaku/analysis/abc/node8.html\#SECTION00870000000000000000}.
We fixed the centers of the Fe K$\alpha$ and K$\beta$ lines at 6.402~keV and 7.060~keV, respectively,
and constrained the K$\beta$ line flux to 12\% of the K$\alpha$ line flux \citep{Thompson2009book}.
We also fixed the hottest plasma temperature at 4.5~keV due to limited photon statistics at high energies.
The best-fit model is very similar to that measured for the \SUZAKU\ data in 2009 \citep{Hamaguchi2014b}.
We included this best-fit model of the CCE and XHN contributions in our analysis of the \NUS\ data near the deep minimum,
with the normalization scaled by a factor of 1.05 to account for the instrumental normalization difference between the 
\SUZAKU\ XIS0 and \NUS/FPMA \citep{Madsen2015a}.

\subsection{First Observation}

\begin{figure}[t]
\plotone{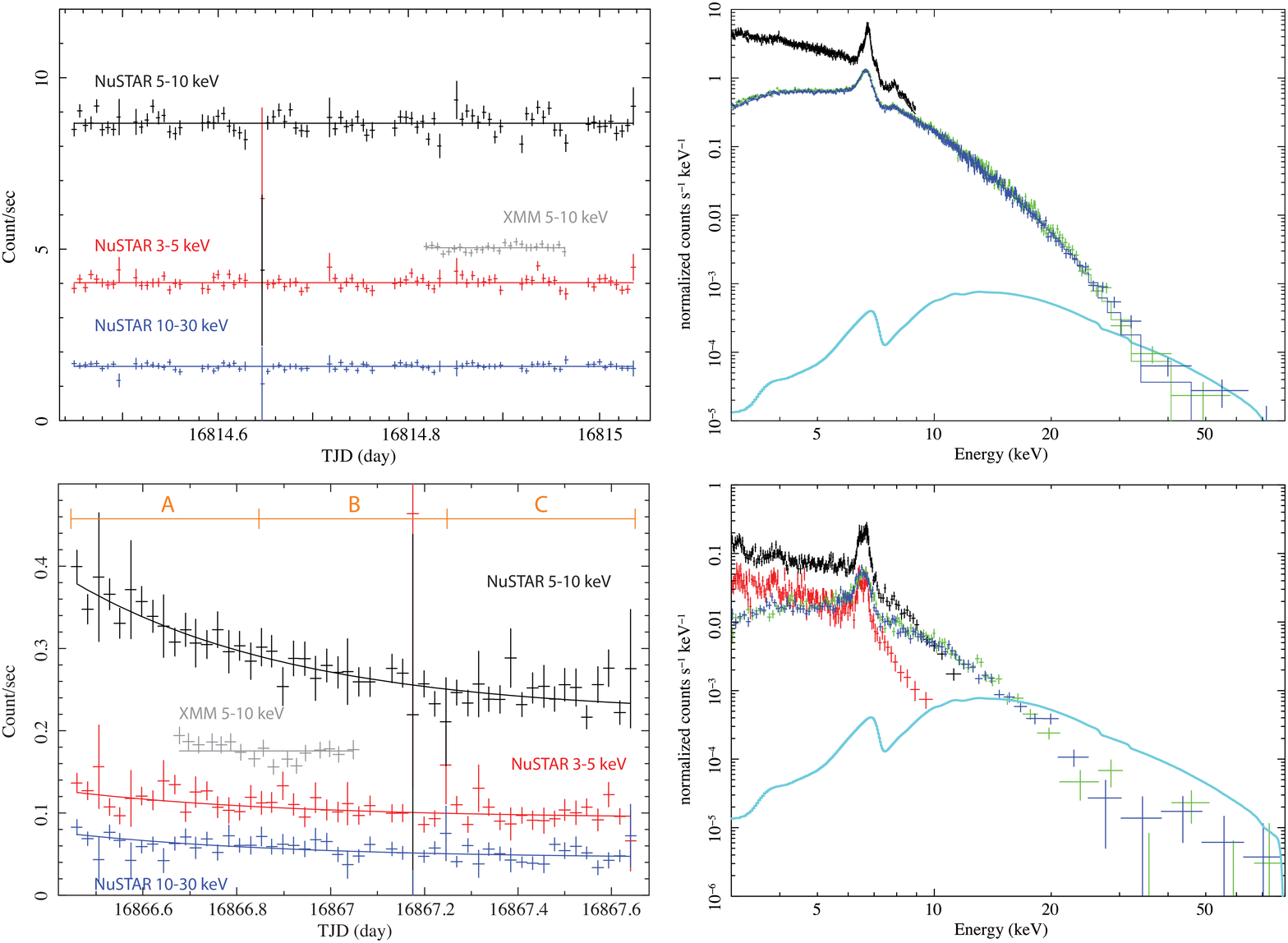}
\caption{
Light curves ({\it left}) and spectra ({\it right}) of the first ({\it top}) and second ({\it bottom}) observations.
{\it Left}: {\XMM}/EPIC-pn ({\it grey}, 5$-$10~keV) and {\NUS}/FPMA+FPMB 
({\it red}: 3$-$5~keV, {\it black}: 5$-$10~keV, {\it blue}: 10$-$30~keV) light curves.
Each light curve bin has 500~sec for the first observation and 2000~sec for the second observation, respectively.
{\it Right}: \XMM/EPIC-pn, MOS1 ({\it black}, {\it red}) and \NUS/FPMA, /FPMB ({\it green, blue}) spectra of \etacar.
The solid lines on the June 6 spectra show the best-fit model in Table~\ref{tbl:spec_bestfit}.
The solid cyan line on each panel for spectra shows the power-law component measured from the \SUZAKU\ observations \citep{Hamaguchi2014b},
convolved with the \NUS/FPMA response.
We do not simultaneously fit the \XMM\ and \NUS\ spectra for July 28 because the \NUS\ spectrum changed significantly 
during the second observation.
\label{fig:obs1st}
\label{fig:obs2nd_lc}
\label{fig:obs2nd_spec}
}
\end{figure}

The \XMM\ observation started 32~ksec after the start of the \NUS\ observation
and covered part of the latter half of the \NUS\ observation (top left panel of Figure~\ref{fig:obs1st}).
During this time, \etacar\ did not show any long-term X-ray variation,
but small flux fluctuations on timescales of $\sim$1~ksec may be present;
the NUS$_{140606}$ light curve between 5$-$10~keV does not accept a constant model at above 3$\sigma$ (reduced $\chi^{2}$ =1.66, d.o.f. =80),
though the light curve appears to be flat.
A flat light curve with possible small fluctuations is typical of \etacar\ \citep{Hamaguchi2007b}.
These small fluctuations may be the low intensity end of the X-ray flares of \etacar\
discussed in detail in \citet{Moffat2009}.

\begin{figure}[t]
\plotone{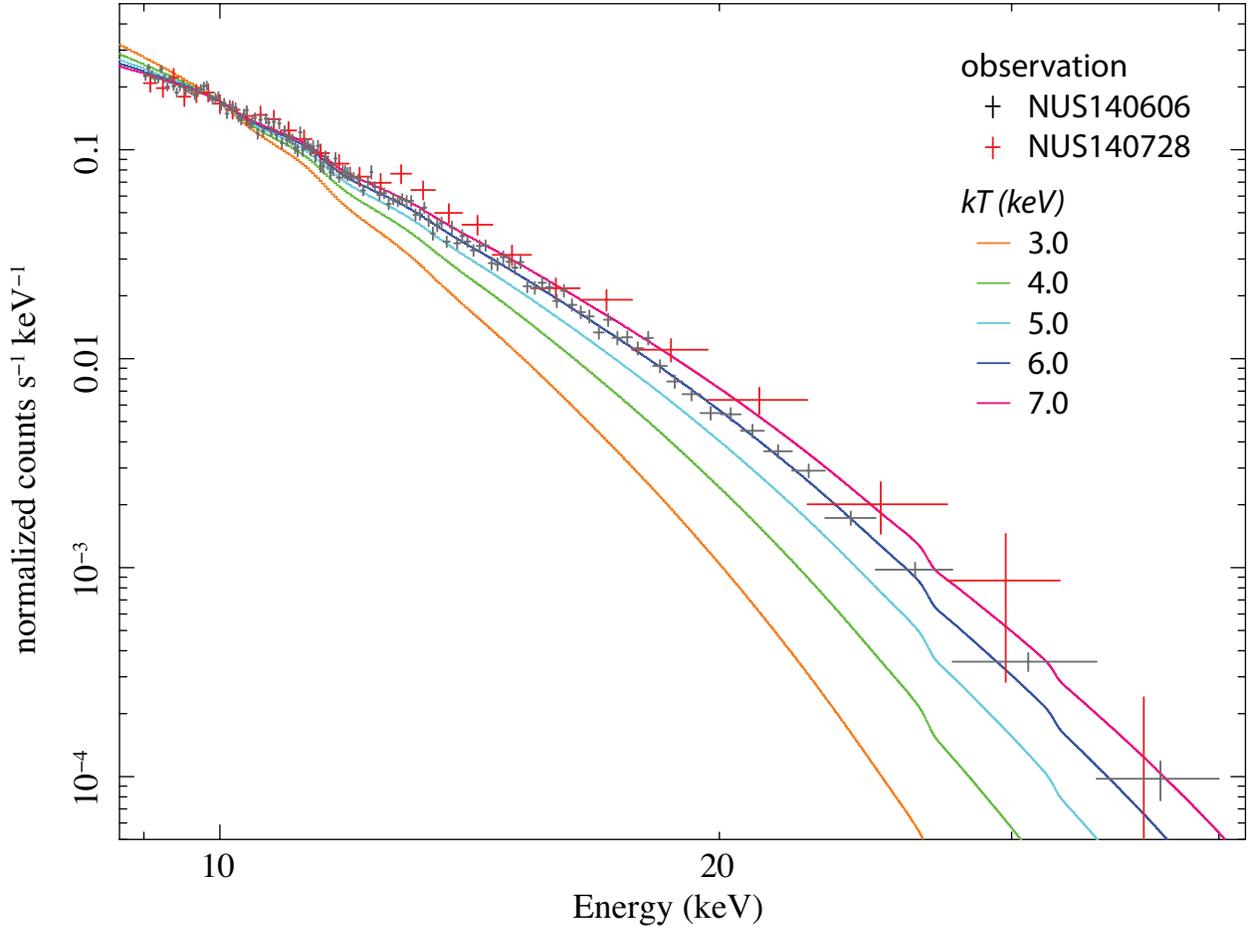}
\caption{
9$-$40~keV spectra of NUS$_{140606}$ ({\it black}) and NUS$_{140728}$ ({\it red}) overlaid.
The NUS$_{140728}$ spectrum is shifted vertically to match the NUS$_{140606}$ spectrum at 10 keV.
The plot also shows bremsstrahlung models at \KT = 3.0, 4.0, 5.0, 6.0 and 7.0, which are normalized at 10~keV, as well.
}
\label{fig:nus_spec_overlaid}
\end{figure}

The top right panel of Figure~\ref{fig:obs1st} shows the \XMM\ and \NUS\ spectra of \etacar\ above 3~keV during these observations.
The \NUS\ spectrum clearly extends up to $\sim$50~keV
and is the first clear detection of the hard thermal tail unambiguously associated with \etacar.
The spectral slope above $\sim$9~keV matches very well with optically-thin thermal emission from 
\KT~$\sim$6~keV plasma (Figure~\ref{fig:nus_spec_overlaid}).
The \XMM\ spectra clearly show emission lines at around 6$-$7~keV, which originate from hydrogen-like, helium-like 
and nearly neutral fluorescent iron ions, as seen in earlier \etacar\ spectra \citep[e.g.,][]{Hamaguchi2007b}.
However, using the nominal detector calibration, these lines were significantly shifted to the blue side by $\sim$40$-$60~eV.
After careful analyses of the emission lines at lower energies, especially compared with results of the Reflection Grating Spectrometers (RGS), and the position of 
the instrumental Au-edge of the mirror coating, we can rule out that the line shifts seen in the EPIC-pn spectrum are due 
to charge transfer inefficiency effects but consistent with a general gain shift. 
Thus we include an additional gain component in our \XMM\ EPIC-pn fits in order to correct for these blue shifts.
It is likely that a flatter \XMM\ spectral slope in the 7$-$10~keV band than {\NUS}'s
is also related to this \XMM\ gain calibration issue.

Both of the \NUS/FPMA \& FPMB spectra show marginal excess above 50~keV over the extrapolation of the thermal tail,
but this excess is smaller than the raw background count rate.
Since the image above 50~keV shows no hint of a point source at the \etacar\ position,
the excess is probably caused by variations in the detector background.
Using Poisson statistics for the background events,
the 3$\sigma$ flux upper-limit between 50$-$70~keV, where the WWC thermal tail drops enough, is 4.0$\times$10$^{-4}$~\UNITCPS~sensor$^{-1}$,
which corresponds to 2.8$\times$10$^{-12}$~\UNITFLUX\ assuming a $\Gamma =$1.4 power-law spectrum.
Regardless of its origin, this excess is below the flux at these energies measured by \INTEGRAL\ and \SUZAKU\ 
(see the solid cyan line in the top right panel of Figure~\ref{fig:obs1st}).

The \SUZAKU\ spectra of \etacar\ obtained between 2005$-$2011
suggest the presence of plasmas in both equilibrium and non-equilibrium conditions \citep{Hamaguchi2014b}.
Since the \XMM\ and \NUS\ spectra do not have enough photon statistics to investigate this feature independently,
we simultaneously fit these spectra by the same spectral model for the \SUZAKU\ spectral fit in \citet{Hamaguchi2014b},
except that we do not include a power-law component.
We freed the model normalizations of \NUS/FPMA and of \NUS/FPMB to the \XMM/EPIC-pn's,
while we fixed the ionization timescale at 7.8$\times$10$^{10}$~cm$^{3}$ s$^{-1}$ ---
the best-fit value of the \SUZAKU\ spectrum in a similar orbital phase in the last cycle ---
because this parameter is less sensitive with free detector gain.
The best-fit result is shown in Table~\ref{tbl:spec_bestfit} and Figure~\ref{fig:obs1st}.
The hottest plasma temperature \KT~$\sim$5.8~keV was significantly higher than the plasma temperature 
measured in earlier observations, which were typically \KT~$\sim$4.5 keV.
The elemental abundance, mainly measured from the iron $K$ emission line fluxes,
is sub-solar and lower than the earlier \SUZAKU\ measurement \citep{Hamaguchi2014b}. 
This is possibly caused by a fit of multi-temperature plasma emission by a simple 2$T$ plasma model.
The other parameters are similar to those from X-ray spectra obtained around the X-ray maximum in 2009.
The spectrum can also be fit by a \KT~$\sim$4.5~keV thermal plasma model plus a hard power-law component
with a similar reduced $\chi^2$ value; however, for this model, 
the power-law index ($\Gamma \sim$4.2) is much steeper than that derived from fits to \INTEGRAL\ and \SUZAKU\ spectra,
and the absorption to the power-law component is unexpectedly high (\NH~$\sim$10$^{24}$~\UNITNH).

\subsection{Second Observation}
\label{subsec:lc_spec_second_obs}

The second \XMM\ observation started 20~ksec after the \NUS\ observation start
and spanned the middle of the \NUS\ observation (see the bottom left panel of Figure~\ref{fig:obs2nd_lc}).
The short \XMM\ observation for $\sim$34~ksec did not show any clear time variation,
but the long \NUS\ observation for $\sim$102~ksec
displayed an obvious flux decrease by $\sim$40\% above $\sim$5~keV.
Such a strong variation has never been seen before in a single pointed observation of \etacar, which is normally very stable on timescales 
of $\lesssim$1~day \citep{Hamaguchi2007b}.
This declining rate is, however, consistent with the average flux decline just before the deep X-ray minimum,
which is measured from the \SWIFT\ monitoring observations (Figure~\ref{fig:variation_minimum}).

The 5$-$10~keV light curve seems to prefer an exponential decay over a constant value.
We therefore modeled this light curve by an exponential plus constant function and found an acceptable 
fit, with an $e$-folding time of 0.48~(0.34$-$0.78)~days, a normalization of 0.12 (0.098$-$0.15)~\UNITCPS\ at 16866.6~day in $TJD$,
and a constant at 0.22~(0.19$-$0.24)~\UNITCPS\ (reduced $\chi^2$ =0.56, d.o.f. =48).
Since this $e$-folding time is roughly consistent with that of the \SWIFT\ light curve before the deep minimum ($\approx$0.9~days),
we suggest that the constant flux component arises from the circumstellar X-ray contamination near \etacar\ (the CCE + XHN emission) 
that is seen clearly only during the deep minimum.
However, the constant flux we derive is a factor of 2 larger than that estimated from 
the best-fit deep minimum spectrum, convolved with the \NUS\ response (0.11~\UNITCPS, see also subsection~\ref{subsec:event_extract}).
A fit of the 5$-$10~keV light curve, fixing the constant at 0.11~\UNITCPS, also gives an acceptable result --- an $e$-folding time of 1.5~(1.4$-$1.7)~day and 
a normalization of 0.23 (0.22$-$0.24)~\UNITCPS\ (reduced $\chi^2$ =0.73, d.o.f. =49).
With this decay rate, the variable emission should be negligible ($\lesssim$10\%) against the stable emission
in $\sim$4.7 days (August 2); this is consistent with the \SWIFT\ light curve, which also suggests the onset of the deep minimum 
around this time (Corcoran et al. in prep).

The 3$-$5~keV and 10$-$30~keV light curves also show flux declines 
though with poorer statistics.
We therefore fixed the $e$-folding time at 0.48~day in their fits and only derived normalizations of the exponential function and 
the constant component.
Compared to the 5$-$10~keV light curve, the 10$-$30~keV light curve has similar contribution from the constant emission,
while the 3$-$5~keV light curve shows a somewhat larger contribution.
This result perhaps suggests a soft X-ray component that does not vary so strongly as the hard X-ray component does.

The bottom right panel of Figure~\ref{fig:obs2nd_spec} shows the \XMM/EPIC-pn, MOS1 and \NUS/FPMA, FPMB spectra
extracted from the entire second observation.
The \XMM\ spectra show two strong peaks around 6$-$7~keV.
The lower energy peak centered at 6.4~keV is the iron fluorescence line, 
while the higher energy peak is the Fe~K thermal emission line complex.
A significant part of the iron fluorescent line should originate from the XHN,
whose reflected emission becomes more prominent as the direct WWC emission declines.
The spectra also show emission lines at 3.9~keV from Ca~K$\alpha$ and at 3.1~keV from Ar~K$\alpha$.
The \NUS\ spectra extend up to $\sim$40~keV.
The spectrum above $\sim$10~keV has a similar slope to that of NUS$_{140606}$,
suggesting the presence of \KT\ $\sim$6~keV plasma (Figure~\ref{fig:nus_spec_overlaid}).
The \NUS\ spectra also show an apparent small excess around 40$-$50~keV, 
but, again, this excess is lower than the background fluctuation, and 
the \NUS\ image above 30~keV does not show any obvious point source at the position of \etacar.
The 3$\sigma$ upper-limit between 40$-$70~keV was 3.5$\times$10$^{-4}$~\UNITCPS~sensor$^{-1}$,
which corresponds to 1.1$\times$10$^{-12}$~\UNITFLUX\ assuming a power-law spectrum
with a photon index consistent with the \INTEGRAL\ and \SUZAKU\ spectra ($\Gamma =$1.4).

\begin{figure}[t]
\plotone{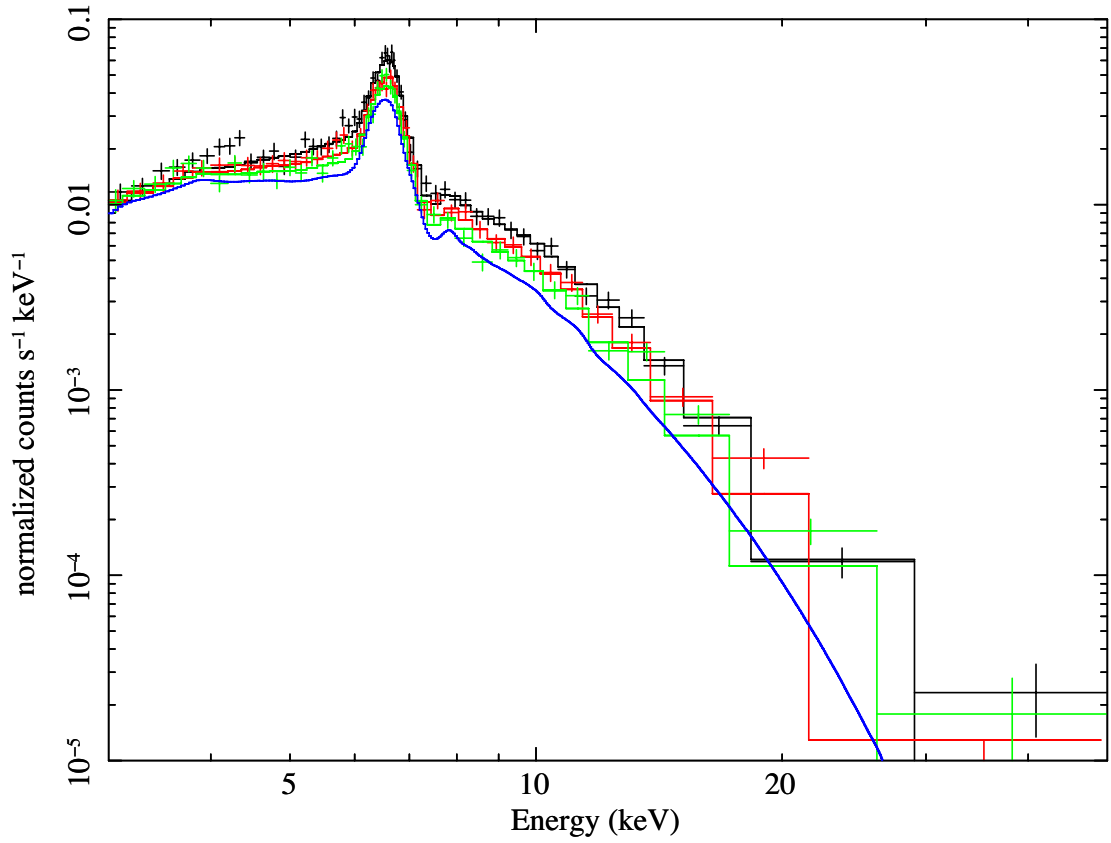}
\caption{
{\NUS}/FPMA+FPMB spectra of the second observation in 3 intervals (A: {\it black}, B; {\it red}, C: {\it green}).
The solid blue line shows the deep minimum spectrum, estimated from the \SUZAKU\ observation on August 6th
and convolved with the \NUS\ response.
\label{fig:obs2nd_spec_split}
}
\end{figure}

We split the \NUS\ observation into three evenly spaced intervals 
(A, B and C: see bottom left of Figure~\ref{fig:obs2nd_lc})
and extracted spectra from each interval to track the spectral variation (Figure~\ref{fig:obs2nd_spec_split}).
The spectral shape above 5~keV did not apparently change between the intervals,
while the spectral normalization decreased.
As seen from the band sliced light curves,
the spectrum below 5~keV is rather unchanged within the photon statistics, suggesting the presence of a relatively stable soft component.
This is similar to the behavior observed in 2009, in which the soft band flux gradually decreased before the onset of the deep X-ray minimum, while the hard band flux dropped sharply \citep[see the middle panel of Figure~2 in][]{Hamaguchi2014a}.

Before performing the spectral fittings, we calibrated the spectral normalizations between instruments.
Since the X-ray flux varied through the \NUS\ observation,
we generated \NUS\ spectra of \etacar\ only during the \XMM\ observation
and simultaneously fit them with the \XMM\ spectra by an empirical model, free of the instrumental normalization ratio.
The results (Table~\ref{tbl:spec_bestfit}) were similar to those measured for XMM$_{140606}$.
We then fit the \XMM\ spectra and the \NUS\ spectra of three intervals simultaneously.
We fixed the instrumental normalization ratios at the values derived above.
We used the same spectral model used to fit the June 6 spectra and tied the physical parameters between the intervals,
except for the normalizations of the WWC component and the fluorescent iron line.
Because of the limited spectral quality,
we fixed the elemental abundance at 1 solar value as derived from the simultaneous fit to the multiple \SUZAKU\ spectra \citep{Hamaguchi2014b}.
The best-fit result is shown at the right column of Table~\ref{tbl:spec_bestfit}.
The absorption to the hard X-ray emission, measured from the iron absorption edge, increased to an extreme value (\NFE~$\sim$9.7$\times$10$^{23}$~\UNITNH)
from the first observation.
This result suggests that very hot plasma at the WWC apex was embedded further into the primary wind.

\section{Discussion}

The plasma temperature in XMM/NUS$_{140606}$, \KT~$\sim$6~keV, was significantly higher than the typical 
plasma temperatures of \etacar\ measured from earlier observations \citep[\KT~$\sim$4$-$5~keV, e.g.][]{Hamaguchi2007b}.
This measurement is weighted strongly by the slope of the bremsstrahlung continuum above 10~keV
in the \NUS\ spectra, while
the flux ratio of the helium-like and hydrogen-like Fe K lines
is still consistent with a more typical temperature, \KT~$\sim$4~keV.
The 6 keV plasma temperature we derive is not perhaps caused by enhanced WWC activity in this cycle
but by stronger contribution of the thermal continuum in the spectral fit.
The second set of observations showed a similarly high plasma temperature (\KT~$\sim$6~keV).
Since \etacar\ had a factor of two flux variation between these observations,
{\etacar}'s WWC activity can thermalize plasma up to $\sim$6~keV until the X-ray minimum onset.

Our analysis of the second observation yielded one of the highest absorption columns ever derived from 
\etacar\ observations (\NFE~$\sim$10$^{24}$~\UNITNH)\footnote{equivalent hydrogen column density in a solar abundance};
the other highest absorptions were observed right after the deep X-ray minimum \citep[\NH~$\sim$10$^{24}$~\UNITNH,][Hamaguchi et al., in preparation]{Hamaguchi2014a}.
This result suggests that the column density to the WWC plasma peaks during the deep minimum and
supports the hypothesis that the deep minimum is mainly caused by an eclipse of the WWC plasma
by an optically thick absorber.

Through the second \NUS\ observation, the hard ($>$5~keV) X-ray emission gradually declined without showing any significant 
spectral change.
A similar variation was seen in the 7$-$10 keV spectral slope in earlier short observations around periastron \citep{Hamaguchi2007b,Hamaguchi2014a}.
Since the decline was smooth,
this indicates that the WWC plasma is perhaps evenly extended and gradually occulted by an optically thick absorber with a relatively sharp boundary.
The current best estimate of the orbital inclination \citep[$i \approx$130$-$145\DEGREE,][]{Madura2012} does not suggest that
the WWC plasma is occulted by the primary stellar body.
This might indicate that colliding wind source might have crossed the WWC contact discontinuity, 
which should have a relatively sharp density change.

\begin{figure}[t]
\plotone{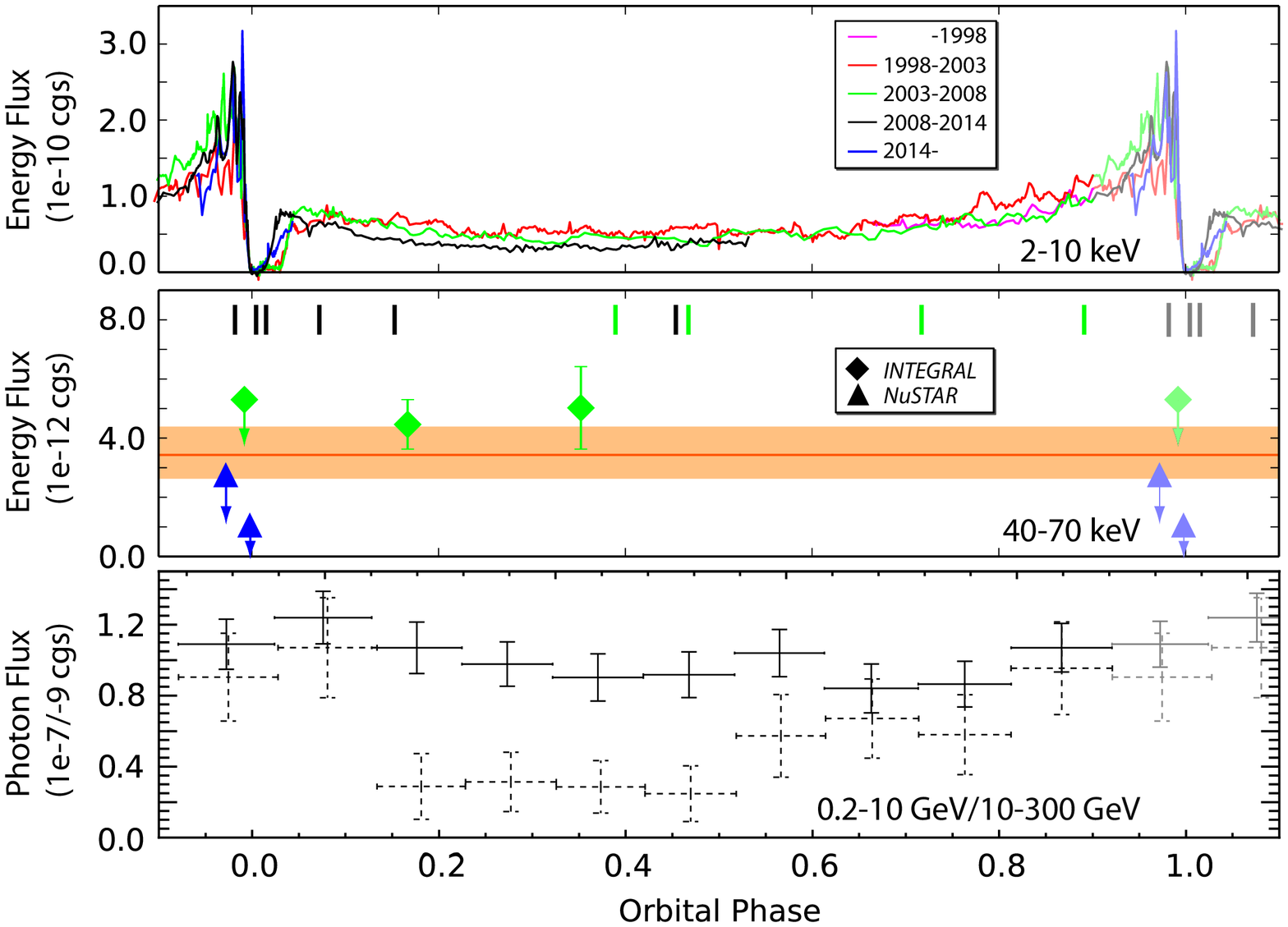}
\caption{
X-ray flux between 2$-$10~keV measured with \RXTE\ and \SWIFT\ \citep[{\it top}:][Corcoran et al. in prep.]{Corcoran2010}
and between 40$-$70~keV measured with \SUZAKU, \INTEGRAL, and \NUS\ \citep[{\it middle}:][]{Leyder2008,Hamaguchi2014b},
and 0.2$-$10~GeV ({\it solid line}) and 10$-$300 GeV ({\it dotted line}) $\gamma$-ray fluxes measured with \FERMI\ \citep[{\it bottom}:][]{Reitberger2015}.
The orange line and shaded area in the middle panel show the best-fit flux and its 90\% error range
of the power-law component, derived from the \SUZAKU\ HXD/PIN spectra below 40~keV assuming a $\Gamma=$1.4 power-law \citep{Hamaguchi2014b}.
The multiple vertical bars on the ceiling of the middle panel show the timings of the \SUZAKU\ observations used for this spectral fit.
\label{fig:flux_nonthermal}
}
\end{figure}

\NUS\ did not detect non-thermal X-ray emission at very high energies.
The upper-limit flux between 40$-$70~keV in NUS$_{140728}$ is 1/4.2 of the \INTEGRAL\ measurement and 1/3.3 of the \SUZAKU\ 
measurement assuming a $\Gamma = 1.4$ power-law spectrum (middle panel of Figure~\ref{fig:flux_nonthermal}).
This result is very surprising because the power-law component was apparently stable between 2004 and 2011.
Interestingly, a \SUZAKU\ observation in 2013 July with a very long exposure of 180~ksec did not detect an excess in the 25$-$40~keV band 
(Yuasa et al. in prep.), so that the power-law source might be variable, and if so it may have decreased before the first \NUS\ observation.

\citet{Reitberger2015} argued that the GeV $\gamma$-ray source was bright through 1 orbital cycle between 2008 August 4 
and 2014 February 18. It appears that this source kept increasing in brighteness through the 2014 periastron, according to the 
1-degree aperture photometry lightcurves weekly created by the \FERMI\ team 
(LAT 3FGL catalog aperture photometry light curves\footnote{http://fermi.gsfc.nasa.gov/ssc/data/access/lat/4yr\_catalog/ap\_lcs.php?ra=10-11, http://fermi.gsfc.nasa.gov/ssc/data/access/lat/4yr\_catalog/ap\_lcs/lightcurve\_3FGLJ1045.1-5941.png}).
This means that the GeV $\gamma$-ray source and the extremely hard (20$-$100~keV) X-ray source behaved
differently around the 2014 periastron passage.
One possible explanation of this discrepancy is that the line of sight column to the $\gamma$-ray source increased before the first \NUS\ observation,
so that extremely hard X-ray emission from the $\gamma$-ray source was totally absorbed.
To suppress the 20$-$70~keV flux by $\lesssim$10\%,
the absorption column should increase to \NH~$\gtrsim$2$\times$10$^{24}$~\UNITNH, 
which can be produced if the $\gamma$-ray source is around the line of sight to the WWC apex.
The other explanation is that the $\gamma$-ray source is unrelated to the hard X-ray source.

\section{Summary}

We performed two simultaneous X-ray observations of \etacar\ with \XMM\ and \NUS\ around the 2014.6 periastron passage.
The {\NUS}'s multi-layer coating mirrors provided the highest spatial resolution observations of extremely hard X-ray emission 
from \etacar.
The simultaneous observations with \XMM, which has good spectral resolution and high sensitivity below $\sim$8~keV,
enabled measurement of the Fe $K$ emission line profile in detail and helped constrain the high-energy thermal tail seen by \NUS.

The \NUS\ and \XMM\ spectra clearly showed that the thermal X-ray slope of \etacar\ extends up to 40$-$50~keV.
This slope is consistent with bremsstrahlung thermal emission from plasma at \KT~$\sim$6~keV,
which was 1$-$2~keV higher than the ionization temperature of Fe K shell ions and
the plasma temperatures measured in earlier observations from spectra below 10~keV.
This slope did not change between the first and second observations though the X-ray flux declined by a factor of 20.
The WWC plasma, or at least a portion of it, did not cool across the X-ray flux decline.

During the second observation, the X-ray flux above 5~keV gradually declined by $\sim$40\% in a day.
This decline is consistent with the deep minimum onset on August 1st and 
can be reproduced with a constant flux plus an exponential decay with an $e$-folding time of 0.5$-$1.5~day.
We did not observe any color variation during the decline, which suggests that the hottest plasma was gradually hidden.
The emission suffered extremely strong absorption (\NFE~$\sim$10$^{24}$~\UNITNH), which is as high as 
the absorption to the WWC plasma right after the deep minimum.
This result supports the hypothesis that the deep minimum is caused by a total eclipse of the WWC apex at superior conjunction.

The \NUS\ data showed no hint of power-law emission above $\sim$30~keV within the \INTEGRAL\ error circle,
giving an upper-limit below the \INTEGRAL\ and \SUZAKU\ detection before 2011.
This indicates that the power-law source probably weakened between the \SUZAKU\ observation in 2011 and the first \NUS\ observation in 2014.
Interestingly, the GeV $\gamma$-ray source seen by \FERMI\ was rather stable around this periastron passage.
This either implies an increase of the absorption to the power-law source during these observations, or that the extremely hard X-ray and GeV $\gamma$-ray sources are unrelated.

\acknowledgments

This research has made use of data obtained from the High Energy Astrophysics Science Archive
Research Center (HEASARC), provided by NASA's Goddard Space Flight Center.
This research has made use of NASA's Astrophysics Data System Bibliographic Services.
We appreciate the \XMM\ help desk and calibration team on helping resolve the \XMM\ EPIC gain issue.
K.H. is supported by the \CHANDRA\ grant GO4-15019A, the \XMM\ grant NNX15AK62G, and 
the ADAP grant NNX15AM96G.
CMPR is supported by an appointment to the NASA Postdoctoral Program at the Goddard Space Flight Center, 
administered by Oak Ridge Associated Universities through a contract with NASA.

Facilities: \facility{XMM (EPIC)}, \facility{NuSTAR}


\begin{thebibliography}{32}
\expandafter\ifx\csname natexlab\endcsname\relax\def\natexlab#1{#1}\fi

\bibitem[{{Abdo} {et~al.}(2010){Abdo}}]{Abdo2010}
{Abdo}, A.~A., {Ackermann}, M., {Ajello}, M., et al. 2010, \apj, 723, 649


\bibitem[{{Aschenbach} {et~al.}(2000){Aschenbach}, {Briel}, {Haberl},
  {Br\"{a}uninger}, {Burkert}, {Oppitz}, {Gondoin}, \& {Lumb}}]{Aschenbach2000}
{Aschenbach}, B., {Briel}, U.~G., {Haberl}, F., et al. 2000, in SPIE,
  Vol. 4012, X-Ray Optics, Instruments, and Missions III, ed. {Joachim E.
  Tr\"{u}mper, Bernd Aschenbach}, p. 731--739

\bibitem[{{Corcoran}(2005)}]{Corcoran2005}
{Corcoran}, M.~F. 2005, \aj, 129, 2018

\bibitem[{{Corcoran} {et~al.}(2004){Corcoran}, {Hamaguchi}, {Gull}, {Davidson},
  {Petre}, {Hillier}, {Smith}, {Damineli}, {Morse}, {Walborn}, {Verner},
  {Collins}, {White}, {Pittard}, {Weis}, {Bomans}, \& {Butt}}]{Corcoran2004}
{Corcoran}, M.~F., {Hamaguchi}, K., {Gull}, T., et al. 2004, \apj, 613, 381

\bibitem[{{Corcoran} {et~al.}(2010){Corcoran}, {Hamaguchi}, {Pittard},
  {Russell}, {Owocki}, {Parkin}, \& {Okazaki}}]{Corcoran2010}
{Corcoran}, M.~F., {Hamaguchi}, K., {Pittard}, J.~M., et al. 2010, \apj, 725, 1528

\bibitem[{{Corcoran} {et~al.}(2001){Corcoran}, {Swank}, {Petre}, {Ishibashi},
  {Davidson}, {Townsley}, {Smith}, {White}, {Viotti}, \&
  {Damineli}}]{Corcoran2001a}
{Corcoran}, M.~F., {Swank}, J.~H., {Petre}, R., et al. 2001, \apj, 562, 1031

\bibitem[{{Damineli} {et~al.}(1997){Damineli}, {Conti}, \&
  {Lopes}}]{Damineli1997}
{Damineli}, A., {Conti}, P.~S., \& {Lopes}, D.~F. 1997, \na, 2, 107

\bibitem[{{Damineli} {et~al.}(2008){Damineli}, {Hillier}, {Corcoran}, {Stahl},
  {Levenhagen}, {Leister}, {Groh}, {Teodoro}, {Albacete Colombo}, {Gonzalez},
  {Arias}, {Levato}, {Grosso}, {Morrell}, {Gamen}, {Wallerstein}, \&
  {Niemela}}]{Damineli2008}
{Damineli}, A., {Hillier}, D.~J., {Corcoran}, M.~F., et al. 2008, \mnras, 384, 1649

\bibitem[{{Farnier} {et~al.}(2011){Farnier}, {Walter}, \&
  {Leyder}}]{Farnier2011}
{Farnier}, C., {Walter}, R., \& {Leyder}, J.-C. 2011, \aap, 526, A57

\bibitem[{{Groh} {et~al.}(2012){Groh}, {Hillier}, {Madura}, \&
  {Weigelt}}]{Groh2012}
{Groh}, J.~H., {Hillier}, D.~J., {Madura}, T.~I., \& {Weigelt}, G. 2012,
  \mnras, 423, 1623

\bibitem[{{Hamaguchi} {et~al.}(2007){Hamaguchi}, {Corcoran}, {Gull},
  {Ishibashi}, {Pittard}, {Hillier}, {Damineli}, {Davidson}, {Nielsen}, \&
  {Kober}}]{Hamaguchi2007b}
{Hamaguchi}, K., {Corcoran}, M.~F., {Gull}, T., et al. 2007, \apj, 663, 522

\bibitem[{{Hamaguchi} {et~al.}(2014{\natexlab{a}}){Hamaguchi}, {Corcoran},
  {Russell}, {Pollock}, {Gull}, {Teodoro}, {Madura}, {Damineli}, \&
  {Pittard}}]{Hamaguchi2014a}
{Hamaguchi}, K., {Corcoran}, M.~F., {Russell}, C.~M.~P., et al. 2014{\natexlab{a}}, \apj, 784, 125

\bibitem[{{Hamaguchi} {et~al.}(2014{\natexlab{b}}){Hamaguchi}, {Corcoran},
  {Takahashi}, {Yuasa}, {Ishida}, {Gull}, {Pittard}, {Russell}, \&
  {Madura}}]{Hamaguchi2014b}
{Hamaguchi}, K., {Corcoran}, M.~F., {Takahashi}, H., et al. 2014{\natexlab{b}}, \apj, 795, 119

\bibitem[{{Harrison} {et~al.}(2013){Harrison}, {Craig}, {Christensen},
  {Hailey}, {Zhang}, {Boggs}, {Stern}, {Cook}, {Forster}, {Giommi},
  {Grefenstette}, {Kim}, {Kitaguchi}, {Koglin}, {Madsen}, {Mao}, {Miyasaka},
  {Mori}, {Perri}, {Pivovaroff}, {Puccetti}, {Rana}, {Westergaard}, {Willis},
  {Zoglauer}, {An}, {Bachetti}, {Barri{\`e}re}, {Bellm}, {Bhalerao},
  {Brejnholt}, {Fuerst}, {Liebe}, {Markwardt}, {Nynka}, {Vogel}, {Walton},
  {Wik}, {Alexander}, {Cominsky}, {Hornschemeier}, {Hornstrup}, {Kaspi},
  {Madejski}, {Matt}, {Molendi}, {Smith}, {Tomsick}, {Ajello}, {Ballantyne},
  {Balokovi{\'c}}, {Barret}, {Bauer}, {Blandford}, {Brandt}, {Brenneman},
  {Chiang}, {Chakrabarty}, {Chenevez}, {Comastri}, {Dufour}, {Elvis}, {Fabian},
  {Farrah}, {Fryer}, {Gotthelf}, {Grindlay}, {Helfand}, {Krivonos}, {Meier},
  {Miller}, {Natalucci}, {Ogle}, {Ofek}, {Ptak}, {Reynolds}, {Rigby},
  {Tagliaferri}, {Thorsett}, {Treister}, \& {Urry}}]{Harrison2013}
{Harrison}, F.~A., {Craig}, W.~W., {Christensen}, F.~E., et al. 2013, \apj, 770, 103

\bibitem[{{Hillier} {et~al.}(2001){Hillier}, {Davidson}, {Ishibashi}, \&
  {Gull}}]{Hillier2001}
{Hillier}, D.~J., {Davidson}, K., {Ishibashi}, K., \& {Gull}, T. 2001, \apj,
  553, 837

\bibitem[{{Leyder} {et~al.}(2008){Leyder}, {Walter}, \& {Rauw}}]{Leyder2008}
{Leyder}, J.-C., {Walter}, R., \& {Rauw}, G. 2008, \aap, 477, L29

\bibitem[{{Leyder} {et~al.}(2010){Leyder}, {Walter}, \& {Rauw}}]{Leyder2010}
---. 2010, \aap, 524, A59

\bibitem[Madsen et al.(2015)]{Madsen2015a} Madsen, K.~K., Harrison, 
F.~A., Markwardt, C.~B., et al.\ 2015, \apjs, 220, 8 

\bibitem[{{Madura} {et~al.}(2012){Madura}, {Gull}, {Owocki}, {Groh}, {Okazaki},
  \& {Russell}}]{Madura2012}
{Madura}, T.~I., {Gull}, T.~R., {Owocki}, S.~P., et al. 2012, \mnras, 420, 2064

\bibitem[{{Mehner} {et~al.}(2010){Mehner}, {Davidson}, {Ferland}, \&
  {Humphreys}}]{Mehner2010b}
{Mehner}, A., {Davidson}, K., {Ferland}, G.~J., \& {Humphreys}, R.~M. 2010,
  \apj, 710, 729

\bibitem[HESS Collaboration et al.(2012)]{HESSCollaboration2012a} HESS 
Collaboration, Abramowski, A., Acero, F., et al.\ 2012, \mnras, 424, 128 

\bibitem[{{Moffat} \& {Corcoran}(2009)}]{Moffat2009}
{Moffat}, A.~F.~J., \& {Corcoran}, M.~F. 2009, \apj, 707, 693

\bibitem[{{Ohm} {et~al.}(2015){Ohm}, {Zabalza}, {Hinton}, \&
  {Parkin}}]{Ohm2015}
{Ohm}, S., {Zabalza}, V., {Hinton}, J.~A., \& {Parkin}, E.~R. 2015, \mnras,
  449, L132

\bibitem[{{Parkin} {et~al.}(2009){Parkin}, {Pittard}, {Corcoran}, {Hamaguchi},
  \& {Stevens}}]{Parkin2009}
{Parkin}, E.~R., {Pittard}, J.~M., {Corcoran}, M.~F., {Hamaguchi}, K., \&
  {Stevens}, I.~R. 2009, \mnras, 394, 1758

\bibitem[{{Pittard} \& {Corcoran}(2002)}]{Pittard2002}
{Pittard}, J.~M., \& {Corcoran}, M.~F. 2002, \aap, 383, 636

\bibitem[{{Pollock} {et~al.}(2005){Pollock}, {Corcoran}, {Stevens}, \&
  {Williams}}]{Pollock2005}
{Pollock}, A.~M.~T., {Corcoran}, M.~F., {Stevens}, I.~R., \& {Williams}, P.~M.
  2005, \apj, 629, 482

\bibitem[Reitberger et al.(2015)]{Reitberger2015} Reitberger, K., Reimer, A., Reimer, O., \& Takahashi, H.\ 2015, \aap, 577, A100 

\bibitem[{{Sekiguchi} {et~al.}(2009){Sekiguchi}, {Tsujimoto}, {Kitamoto},
  {Ishida}, {Hamaguchi}, {Mori}, \& {Tsuboi}}]{Sekiguchi2009}
{Sekiguchi}, A., {Tsujimoto}, M., {Kitamoto}, S., et al. 2009, \pasj, 61, 629

\bibitem[{{Skinner} {et~al.}(2001){Skinner}, {G{\"u}del}, {Schmutz}, \&
  {Stevens}}]{Skinner2001}
{Skinner}, S.~L., {G{\"u}del}, M., {Schmutz}, W., \& {Stevens}, I.~R. 2001,
  \apjl, 558, L113

\bibitem[{{Smith}(2006)}]{Smith2006b}
{Smith}, N. 2006, \apj, 644, 1151

\bibitem[{{Str{\" u}der} {et~al.}(2001){Str{\" u}der}, {Briel}, {Dennerl},
  {Hartmann}, {Kendziorra}, {Meidinger}, {Pfeffermann}, {Reppin}, {Aschenbach},
  {Bornemann}, {Br{\" a}uninger}, {Burkert}, {Elender}, {Freyberg}, {Haberl},
  {Hartner}, {Heuschmann}, {Hippmann}, {Kastelic}, {Kemmer}, {Kettenring},
  {Kink}, {Krause}, {M{\" u}ller}, {Oppitz}, {Pietsch}, {Popp}, {Predehl},
  {Read}, {Stephan}, {St{\" o}tter}, {Tr{\" u}mper}, {Holl}, {Kemmer},
  {Soltau}, {St{\" o}tter}, {Weber}, {Weichert}, {von Zanthier},
  {Carathanassis}, {Lutz}, {Richter}, {Solc}, {B{\" o}ttcher}, {Kuster},
  {Staubert}, {Abbey}, {Holland}, {Turner}, {Balasini}, {Bignami}, {La
  Palombara}, {Villa}, {Buttler}, {Gianini}, {Lain{\' e}}, {Lumb}, \&
  {Dhez}}]{Struder2001}
{Str{\" u}der}, L., {Briel}, U., {Dennerl}, K., et al. 2001, \aap, 365, L18

\bibitem[{{Tavani} {et~al.}(2009){Tavani}, {Sabatini}, {Pian}, {Bulgarelli},
  {Caraveo}, {Viotti}, {Corcoran}, {Giuliani}, {Pittori}, {Verrecchia},
  {Vercellone}, {Mereghetti}, {Argan}, {Barbiellini}, {Boffelli}, {Cattaneo},
  {Chen}, {Cocco}, {D'Ammando}, {Costa}, {DeParis}, {Del Monte}, {di Cocco},
  {Donnarumma}, {Evangelista}, {Ferrari}, {Feroci}, {Fiorini}, {Froysland},
  {Fuschino}, {Galli}, {Gianotti}, {Labanti}, {Lapshov}, {Lazzarotto},
  {Lipari}, {Longo}, {Marisaldi}, {Mastropietro}, {Morelli}, {Moretti},
  {Morselli}, {Pacciani}, {Pellizzoni}, {Perotti}, {Piano}, {Picozza}, {Pilia},
  {Porrovecchio}, {Pucella}, {Prest}, {Rapisarda}, {Rappoldi}, {Rubini},
  {Soffitta}, {Trifoglio}, {Trois}, {Vallazza}, {Vittorini}, {Zambra},
  {Zanello}, {Santolamazza}, {Giommi}, {Colafrancesco}, {Antonelli}, \&
  {Salotti}}]{Tavani2009}
{Tavani}, M., {Sabatini}, S., {Pian}, E., et al. 2009, \apjl, 698, L142

\bibitem[{{Thompson} {et~al.}(2009){Thompson}, {Lindau}, {Attwood}, {Liu},
  {Gullikson}, {Pianetta}, {Howells}, {Robinson}, {Kim}, {Scofield}, {Kirz},
  {Underwood}, {Kortright}, {Williams}, \& {Winick}}]{Thompson2009book}
{Thompson}, A., {Lindau}, I., {Attwood}, D., et al. 2009, {X-ray Data Booklet} (Center for X-ray optics and advanced light
  source, Lawrence Berkeley National Laboratory, University of California)

\bibitem[{{Turner} {et~al.}(2001){Turner}, {Abbey}, {Arnaud}, {Balasini},
  {Barbera}, {Belsole}, {Bennie}, {Bernard}, {Bignami}, {Boer}, {Briel},
  {Butler}, {Cara}, {Chabaud}, {Cole}, {Collura}, {Conte}, {Cros}, {Denby},
  {Dhez}, {Di Coco}, {Dowson}, {Ferrando}, {Ghizzardi}, {Gianotti}, {Goodall},
  {Gretton}, {Griffiths}, {Hainaut}, {Hochedez}, {Holland}, {Jourdain},
  {Kendziorra}, {Lagostina}, {Laine}, {La Palombara}, {Lortholary}, {Lumb},
  {Marty}, {Molendi}, {Pigot}, {Poindron}, {Pounds}, {Reeves}, {Reppin},
  {Rothenflug}, {Salvetat}, {Sauvageot}, {Schmitt}, {Sembay}, {Short},
  {Spragg}, {Stephen}, {Str{\" u}der}, {Tiengo}, {Trifoglio}, {Tr{\" u}mper},
  {Vercellone}, {Vigroux}, {Villa}, {Ward}, {Whitehead}, \&
  {Zonca}}]{Turner2001}
{Turner}, M.~J.~L., {Abbey}, A., {Arnaud}, M., et al. 2001, \aap, 365, L27

\bibitem[{{Verner} {et~al.}(2005){Verner}, {Bruhweiler}, \&
  {Gull}}]{VernerE2005a}
{Verner}, E., {Bruhweiler}, F., \& {Gull}, T. 2005, \apj, 624, 973

\bibitem[{{Zhekov} \& {Park}(2010)}]{Zhekov2010}
{Zhekov}, S.~A., \& {Park}, S. 2010, \apj, 721, 518

\end{thebibliography}

\begin{deluxetable}{lcccccc}
\tablecolumns{7}
\tablewidth{0pc}
\tabletypesize{\scriptsize}
\tablecaption{Logs of the \XMM\ and \NUS\ Observations\label{tbl:obslogs}}
\tablehead{
\colhead{Observatory}&
\colhead{Abbreviation}&
\colhead{Observation ID}&
\colhead{Observation Start}&
\colhead{$\phi_{\rm X}$}&
\colhead{Duration}&
\colhead{Exposure}\\
&&&&&\colhead{(ksec)}&\colhead{(ksec)}
}
\startdata
\multicolumn{7}{l}{First (Maximum):}\\
~\XMM &XMM$_{140606}$&0742850301&2014 June 6, 19:30 (m1)&2.9721&12.8/13.0&9.0/12.6\\ 
~\NUS&NUS$_{140606}$&30002040002&2014 June 6, 10:31&2.9721&50.6&32.9\\
\multicolumn{7}{l}{Second (Before Minimum):}\\
~\XMM&XMM$_{140728}$&0742850401&2014 July 28, 15:50 (m1)&2.9978&33.5/33.7&23.5/32.6\\
~\NUS&NUS$_{140728}$&30002040004&2014 July 28, 10:31&2.9979&102.1&61.3\\
\multicolumn{7}{l}{Supplement (Deep Minimum):}\\
~\SUZAKU&SUZ$_{140806}$&409028010&2014 August 6, 20:04&3.0025&71.9&21.5\\
\enddata
\tablecomments{
Abbreviation: Abbreviation adopted for each observation. 
Observation ID: Observation identification number of each observation.
Observation Start: Time of the observation start.
$\phi_{\rm X}$: Phase at the center of the observation in the X-ray ephemeris in \citet{Corcoran2005}, $\phi_{\rm X}$ = (JD[observation start] $-$ 2450799.792)/2024.
Duration: Duration of the Observation. 
Exposure: Exposure time excluding the detector deadtime.
For \XMM, the two numbers divided by slash are of EPIC-pn and MOS1, respectively.
}
\end{deluxetable}

\begin{deluxetable}{lcrlrl}
\tablecolumns{6}
\tablewidth{0pc}
\tabletypesize{\scriptsize}
\tablecaption{Best-fit Spectral Model \label{tbl:spec_bestfit}}
\tablehead{
\colhead{Parameter}&
\colhead{Unit}&
\multicolumn{2}{c}{First Observation}&
\multicolumn{2}{c}{Second Observation}
}
\startdata
\multicolumn{6}{l}{Hot Component}\\
~~\KT& 		[keV]			&5.8 &	(5.7$-$5.8)&	5.7 	& (4.8$-$6.4) \\
~~$Z$& 		[solar]			&0.69 &	(0.67$-$0.71)&	1.0 & (fix)\\
~~$\tau$ [{\tt nei}]& [cm$^{3}$ s$^{-1}$] 	&7.8e10&(fixed)$^{\dagger1}$	&	2.0e11 	&(1.5e11$-$2.8e11)\\
~~norm [{\tt nei}]& [cm$^{-5}$]		&3.2e-4&($<$2.0e-3)  &	6.5e-3	&(5.1e-3$-$9.3e-3)\\
~~norm [{\tt apec}]& [cm$^{-5}$]				&0.23 &	(0.22$-$0.23)&	3.0e-14 &($<$5.4e-4)\\
~~norm ratio&\\
~~~~A&			&\nodata	&	& 1.16	&(1.03$-$1.30)\\
~~~~B&			&\nodata	&	& 0.65	&(0.57$-$0.74)\\
~~~~C&			&\nodata	&	& 0.35	&(0.27$-$0.43)\\
~~Gaussian$_{6.4}$ flux&[10$^{-5}$~\UNITCPS] 				&49&(48$-$53)	&	1.8	&(1.1$-$2.5)\\ 
~~~~A&	&\nodata	&	& 1.9		&(0.39$-$3.4)\\
~~~~B&	&\nodata	&	& 0.0		&($<$0.86)\\
~~~~C&	&\nodata	&	& 0.23		&($<$1.5)\\
~~\NH & [10$^{23}$~\UNITNH] 		& 4.2&(4.1$-$4.2)	&	5.4		&(4.2$-$6.7)\\
~~\NFE & [10$^{23}$~\UNITNH] 		& 3.0&(2.8$-$3.1)	&	9.7		&(7.8$-$11.7)\\
\multicolumn{6}{l}{Cool Component}\\
~~\KT& [keV] 				& 2.7	&(2.7$-$2.7)&	1.4		&($>$0.35)\\
~~$Z$& [solar] 					&0.51 &(0.49$-$0.52)&	1.0 & (fixed)\\
~~norm&	[cm$^{-5}$]		&0.34&(0.34$-$0.35)	&	9.2e-4 &(2.6e-4$-$0.13)\\
~~\NH & [\UNITNH]				&5.4 &(5.3$-$5.5)	&	5.0 &(fixed)\\ \hline
\multicolumn{6}{l}{Instrument Normalization}\\
\multicolumn{2}{l}{~\XMM/MOS}	&\nodata&	&0.957$^{\dagger2}$&(0.924$-$0.991)\\
\multicolumn{2}{l}{~\NUS/FPMA}	&1.120& (1.114$-$1.125)			&1.101$^{\dagger2}$& (1.058$-$1.143)\\
\multicolumn{2}{l}{~\NUS/FPMB}	&1.146& (1.141$-$1.151)			&1.127$^{\dagger2}$& (1.083$-$1.172)\\ \hline
\multicolumn{6}{l}{XMM pn gain}\\
~linear &		&1.011&	&\nodata&\\ \hline
reduced $\chi^{2}$ (d.o.f.)&&\multicolumn{2}{c}{1.454 (1117)}&\multicolumn{2}{c}{1.094 (836)}\\
\enddata
\tablecomments{
Model: ({\tt apec}[\KT$_{\rm var}$, $Z_{\rm var}$, norm$_{\rm var}$[{\tt apec}]] + {\tt nei}[\KT$_{\rm var}$, $Z_{\rm var}$, $\tau$[{\tt nei}], norm$_{\rm var}$[{\tt nei}]] + {\tt Gaussian}$_{6.4}$[flux$_{\rm var}$] + {\tt Gaussian}$_{7.1}$[0.12$\times$flux$_{\rm var}$]) {\tt varabs}[\NH$_{\rm var}$, \NFE$_{\rm var}$] + {\tt apec}[\KT$_{\rm const}$, $Z_{\rm const}$, norm$_{\rm const}$] {\tt TBabs}[\NH$_{\rm const}$] + ``the deep minimum spectrum".
The narrow Gaussian components, Gaussian$_{6.4}$ and Gaussian$_{7.1}$, are for the fluorescent Fe K$_{\alpha}$ and K$_{\beta}$ lines, 
and their line center energies are fixed at 6.402~keV and 7.060~keV, respectively.
The Fe K$_{\beta}$ line flux is tied to 12\% of the Fe K$_{\alpha}$ flux \citep{Thompson2009book}.
We assume an independent elemental abundance for the cool component to 
simply reproduce the spectral shape.
In the second observation column, the normalization ratios and the Gaussian$_{6.4}$ fluxes of the A, B and C intervals are obtained from the \NUS\ spectra,
while the other independent parameters (norm$_{\rm var}$[{\tt nei}] and norm$_{\rm var}$[{\tt apec}], Gaussian$_{6.4}$ flux) 
are from the \XMM/EPIC-pn and MOS1 spectra obtained during the \XMM\ observation interval.
The ratios between norm$_{\rm var}$[{\tt apec}] and norm$_{\rm var}$[{\tt nei}] for these spectra are tied together.
Their errors are estimated after fixing the norm$_{\rm var}$[{\tt apec}] and norm$_{\rm var}$[{\tt nei}] parameters 
for the \XMM\ spectra at the best-fit values.
The parentheses quote the 90\% confidence ranges.
$^{\dagger1}$The spectrum is not sensitive to the ionization timescale because of the gain fit.
We therefore fixed it to that of the \SUZAKU\ measurement in a similar orbital phase in the last cycle.
$^{\dagger2}$The best-fit values and errors are measured from a simultaneous fit to the \XMM\ and \NUS\ spectra 
during the \XMM\ observation interval. 
These numbers are fixed in a spectral fit for the whole second observation, and therefore do not affect the fitting result of the other parameters.
}
\end{deluxetable}

\clearpage

\end{document}